\documentclass[manuscript,screen]{acmart}

\AtBeginDocument{%
  \providecommand\BibTeX{{%
    \normalfont B\kern-0.5em{\scshape i\kern-0.25em b}\kern-0.8em\TeX}}}

\setcopyright{acmlicensed}
\copyrightyear{2024}
\acmYear{2024}
\acmDOI{XXXXXXX.XXXXXXX}

\acmConference[FAccT ’24]{Make sure to enter the correct
  conference title from your rights confirmation emai}{June 03--06,
  2024}{Rio de Janeiro, Brazil}
 \acmBooktitle{ACM FAccT Conference 2024}
\acmISBN{978-1-4503-XXXX-X/18/06}

\copyrightyear{2024} 
\acmYear{2024} 
\setcopyright{rightsretained} 
\acmConference[FAccT '24]{The 2024 ACM Conference on Fairness, Accountability, and Transparency}{June 3--6, 2024}{Rio de Janeiro, Brazil}
\acmBooktitle{The 2024 ACM Conference on Fairness, Accountability, and Transparency (FAccT '24), June 3--6, 2024, Rio de Janeiro, Brazil}\acmDOI{10.1145/3630106.3658900}
\acmISBN{979-8-4007-0450-5/24/06}

\begin{document}

\title{A Framework for Exploring the Consequences of AI-Mediated Enterprise Knowledge Access and Identifying Risks to Workers}
\renewcommand{\shorttitle}{Risks and Consequences of AI-Mediated EKA}


\author{Anna Gausen}
\affiliation{%
  \institution{Imperial College London}
  \streetaddress{Exhibition Road}
  \city{London}
  \country{UK}}
\email{anna.gausen16@imperial.ac.uk}

\author{Bhaskar Mitra}
\affiliation{%
  \institution{Microsoft Research}
  \city{Montreal}
  \country{Canada}
}

\author{Siân Lindley}
\affiliation{%
  \institution{Microsoft Research}
  \city{Cambridge}
  \country{UK}
}

\renewcommand{\shortauthors}{Gausen et al.}

\begin{abstract}
Organisations generate vast amounts of information, which has resulted in a long-term research effort into knowledge access systems for enterprise settings. Recent developments in artificial intelligence, in relation to large language models, are poised to have significant impact on knowledge access. This has the potential to shape the workplace and knowledge in new and unanticipated ways. Many risks can arise from the deployment of these types of AI systems, due to interactions between the technical system and organisational power dynamics. 

This paper presents the Consequences-Mechanisms-Risks framework to identify risks to workers from AI-mediated enterprise knowledge access systems. We have drawn on  wide-ranging literature detailing risks to workers, and categorised risks as being to worker value, power, and wellbeing. The contribution of our framework is to additionally consider (i) the consequences of these systems that are of moral import: commodification, appropriation, concentration of power, and marginalisation, and (ii) the mechanisms, which represent how these consequences may take effect in the system. The mechanisms are a means of contextualising risk within specific system processes, which is critical for mitigation. This framework is aimed at helping practitioners involved in the design and deployment of AI-mediated knowledge access systems to consider the risks introduced to workers, identify the precise system mechanisms that introduce those risks, and begin to approach mitigation. Future work could apply this framework to other technological systems to promote the protection of workers and other groups.
\end{abstract}




\received[accepted]{30 March 2024}


\maketitle

\section{Introduction}

People are increasingly interacting with, and being affected by, the deployment of AI systems in the workplace. How AI systems are impacting workers is a pressing matter for system designers, policy-makers, and workers themselves \cite{Myers2023Algorithmic}. This paper will focus on how AI-based systems for enterprise knowledge access (EKA) can specifically impact workers. We propose a framework for identifying and understanding mechanisms that lead to systemic consequences of moral import with the goal of safeguarding worker power, value, and wellbeing.

Organisations generate huge amounts of information that raise challenges associated with the maintenance, dissemination, and discovery of organisational knowledge \cite{larsen-ledet2022ethical}. This has led to a long-term research effort into how knowledge access systems can be developed for enterprise settings. Recent examples of these systems can automatically extract knowledge that an organisation has produced and store it in a knowledge base, from where it can be surfaced to end-users \cite{winn2018alexandria} \cite{vivatopics}. Recent developments in artificial intelligence (AI), notably large language models (LLMs) like OpenAI’s Generative Pre-trained Transformer (GPT4) \cite{openai2023gpt4}, present a shift in what is possible in this domain. These models are trained on exceptionally large datasets and can be applied to a broad set of downstream tasks \cite{Tamkin2021}. Their capabilities could enable more extensive mining, knowledge synthesis, and natural language interaction in relation to knowledge. They could make EKA systems an integral part of people's work, by surfacing knowledge in new ways, including through interactions that are implicit and proactive \cite{ju2008design} \cite{serim2019explicating}.

At this moment, where AI-mediated EKA is enabled in commercially available enterprise technologies, and where this may be expedited and transformed by LLMs, there is a need to consider implications for workers. For the benefits of EKA to be fully realised, worker adoption, organisational responsibility, and high-quality knowledge sharing are all necessary. All three require the system to be designed in a way that minimises negative outcomes for workers. This necessitates looking beyond concerns relating to privacy and surveillance \cite{Akhtar2021} \cite{Abril2023}, and additionally considering the many other risks that can arise from the deployment of these types of AI systems due to the ``co-productive'' \cite{Jasanoff2004}  \cite{Shelby2023} interactions between the technical system and organisational power dynamics \cite{Ajunwa2023}. The research and engineering community is missing a structured and actionable framework for critically examining the consequences of AI-mediated EKA systems in use and identifying potential risks to workers from these systems. 

In this paper, we propose a framework to identify the risks to workers associated with deploying AI-mediated EKA systems. We call this the \textit{Consequences-Mechanisms-Risks Framework}. The framework is aimed at supporting those involved in the design and/or deployment of such systems to identify the risks they introduce, the specific system mechanisms that introduce those risks, and the actionable levers to reduce those risks. Many existing papers identify risks. The contribution of our framework is to also consider consequences, which are systemic changes affected by the system deployment which may correspond to specific risks, and the mechanisms, which cause these consequences and may contribute to certain risks. The mechanisms also help contextualise risk within specific system processes.
As the technical system cannot be separated from organisational socio-cultural dynamics \cite{Alkhatib2021} \cite{Constana-Chock2020}, we consider potential risks to be sociotechnical and system-level rather than by adopting a model-centric approach \cite{weidinger2023sociotechnical}.

This work will have the following research contributions: 

\begin{itemize}
    \item We develop the Consequences-Mechanisms-Risks Framework to explore the consequences of moral import of a system, the system mechanisms that introduce risk, and the risk manifestations. 
    \item We apply the Consequences-Mechanisms-Risks Framework to AI-mediated EKA systems and identify a set of consequences, mechanisms, and risks to workers.
    \item We highlight certain considerations to help practitioners, involved in the design and deployment of such systems, to reduce risk to workers.
\end{itemize}

\section{Literature Survey}  \label{sect - survey}

\subsection{Knowledge and the Enterprise}

The study of knowledge in the enterprise has a long history in the fields of human-computer interaction (HCI), computer-supported cooperative work, organisational studies, and more \cite{Lindley2023}. Organisational knowledge entails both knowledge that is generated through work (i.e., the outputs of ``knowledge work'') as well as the ``know how'' that enables people to get their work done, be this by following workplace policies or by adhering to ``unofficial'' codes of behaviour \cite{phan2000knowledge}, and knowing how to work with and draw on the expertise of others \cite{randall2001memories}. Scholars have differentiated between tacit, implicit, and explicit knowledge, which differ in terms of whether knowledge has been, or can be, articulated, as well as between declarative and procedural knowledge, which can also be framed as the difference between ``know about'' and ``know how'' \cite{nonaka1994dynamic}. Prior research has highlighted that knowledge development and dissemination are inherently social processes \cite{nonaka1994dynamic} \cite{nonaka2000seci}  \cite{nonaka2015knowledge} that are dependent on shared norms, understanding, and connections \cite{ackerman2003sharing}. 

Our paper focuses on AI-mediated EKA specifically. EKA systems include knowledge management systems \cite{schmid2019using} \cite{winn2018alexandria}, enterprise search systems \cite{cleverley2019enterprise}, and enterprise chat interfaces powered by conversational agents. Examples include Amazon’s Kendra, which uses generative AI for conversational agents that sit on top of enterprise content \cite{AmazonKendra}, and Microsoft’s VIVA Topics, which uses AI to automatically compile and surface topic-based information to workers \cite{vivatopics}. Research has explored the potential for risks to arise from the deployment of knowledge systems in enterprise settings. Examples include a discussion of social and ethical considerations relevant to expert identification and  recommendation via workplace technologies \cite{larsen-ledet2022ethical}, and considerations for the responsible design of enterprise knowledge bases produced through implicit interactions between organisation members and technologies \cite{Lindley2023}. AI-mediation can both introduce new challenges, such as system opacity which may make recourse harder, and worsen existing challenges, such as reproducing the historic biases within an organisation. AI-mediated EKA systems have a unique set of challenges, compared to other EKA systems, due to the workplace power asymmetries they are deployed under. This paper advances prior work through the development of a formalised framework to address risk to workers from AI-mediated EKA.

\subsection{Risk to Workers} \label{subsect - survey of risks}

This section will survey literature on risks and harms to workers, with a particular focus on technological systems. Our perspective on risks to workers has been strengthened by wide ranging epistemologies including feminism \cite{Haraway1998} \cite{Wajcman1991} \cite{Johnson2020}, post-colonial theory \cite{Couldry2018} \cite{Thatcher2016}  \cite{Mohamed2020}, labour rights \cite{Prassl2018Humans} \cite{Newman2016}, and design justice \cite{Constana-Chock2020}. We identified three high-level risk areas: (1) reduced worker value, (2) reduced worker power, and (3) reduced worker wellbeing. These risk areas are discussed in detail below.

The first identified risk area in the literature is \textit{reduced worker value}, both perceived and economic \cite{Berkley2022Technological}. Automation can shift economic benefit from workers to the organisation depending on who claims the surplus from efficiency gains \cite{Buolamwini2018}. This is not an inevitable outcome of automation, it could be designed in ways that shares the benefit with workers. Systems can also lead to automation bias, where there is over-reliance on a system. Systems that act as mediators between people ``rearrange social contexts'' \cite{Baym2023} by controlling who and what is seen \cite{larsen2022ways} \cite{lorenzo2023through}. Systems can heighten favourable visibility, heighten unfavourable visibility or lower visibility. This can cause allocative \cite{Dubal2023} \cite{Bernhardt2021} or representative harms. The use of workplace technologies can also result in an environment where workers are treated as ``fungible human capital'' \cite{ajunwa2019platforms}, experience deskilling \cite{Rodrigues2020}, and are seen as easily replaceable. Deskilling of workers may be accompanied by an expectation to reskill. This represents a shifting of power away from workers and puts the burden on them to adapt in the face of changes they have no power over.

The second identified risk area in the literature is \textit{reduced worker power}. AI systems often reflect, reinforce, and codify the inequities built into the power structures into which they are deployed \cite{Benjamin}. Due to the power asymmetries that already exist in the workplace, workers will be particularly vulnerable to this \cite{Ajunwa2023} \cite{Crooks2021}.  Loss of agency happens when an AI-system results in the reduction of autonomy and decision-making power for users \cite{Shelby2023}. Systems that mediate interactions and knowledge sharing may lower inter-worker relations, reduce their social capital \cite{Baym2021WFH}, and cause alienation, impacting their ability to organise and engage in collective action. Collective action is a fundamental defence for workers \cite{Hardt2023} \cite{Rodrigues2020}, enabling them to negotiate other working conditions such as wages or privacy concerns \cite{Newman2016}. Data leverage, the influence individuals or groups have due to the reliance of computing technologies on their data contributions, is one way workers could bargain with the aid of such systems \cite{VincentDataLeverage}.  

The final identified risk area is \textit{reduced worker wellbeing}. This can occur through increased workloads, feelings of surveillance, and less meaningful work. A system may create "work intensification" \cite{Bernhardt2021} if workers are required to maintain and feed data to the system in addition to their existing responsibilities \cite{larsen-ledet2022ethical}. Meaningful work is a key part of Rawlsian Justice in the workplace \cite{Ajunwa2020}, yet a survey of the last century of automation suggests it has reduced meaningful tasks \cite{Crawford2023podcast} \cite{siddarth2021ai}. Workplace systems can result in the unwanted collection of personal data \cite{Ramesh2022} and the collection of data without informed consent \cite{Chowdhary2023}. Data collection can take the form of relying on workers to provide the system with continuous data, a form of participatory surveillance \cite{Ajunwa2017}, or it could mine worker data without their awareness, a form of discreet surveillance \cite{Berkley2022Technological}. This can lead to harms including loss of desired anonymity, non-consensual data collection, loss of the right to be forgotten, non-consensual representation or classification of individuals, and data used for unexpected purposes \cite{Shelby2023}. These three identified risk areas interact: reduction of worker value reduces worker power as it diminishes their bargaining power. The diminished power will, in turn, lower worker value and wellbeing, as it reduces their ability to negotiate better pay or working conditions.

Structured frameworks, including taxonomies and landscape analyses, encourage practitioners to anticipate risks, and formalise social and ethical considerations around deployment. Structured frameworks have been developed to analyse harms, risks, and failures associated with AI and other technological systems, from different perspectives. These frameworks can be oriented by specific types of models \cite{Kirk2023} \cite{Weidinger2022} \cite{Tamkin2021} \cite{bommasani2021opportunities}, by particular harm types \cite{Katzman2023}  \cite{Tran2020} \cite{agrafiotis2018taxonomy}, by system failures \cite{Bandy2021} \cite{Raji2022} \cite{Shen2021} or by domain of use \cite{Scheuerman2021} \cite{Banko2020} \cite{Walker2020} \cite{Tran2020} \cite{Katzman2023} \cite{Angwin2016}. Our analysis will be oriented by domain, focusing on risks to workers from AI-mediated EKA. We have not chosen a model-oriented approach because the risks arise based on how the model is entangled with the socio-technical environment that it operates within. Similarly, we have not narrowed our analysis to a specific type of risk, as many different types of risk can arise from a system of interest. Finally, we are interested in risks that result from functioning systems, which may be unintentional or unanticipated. System failures are out of the scope of this analysis. 

\section{Consequences-Mechanisms-Risks Framework} \label{sect - mechanisms}

The aim of this research is to formulate an actionable framework to help practitioners identify and understand risks to workers associated with the deployment of AI-mediated EKA systems. In this paper, we use the term \textit{AI-mediated enterprise knowledge access (EKA)} to describe a system that has predictive capabilities (AI), and that extracts and surfaces knowledge (knowledge access) used in an enterprise setting, such as retrieval systems or chat interfaces.  This analysis consists of identifying the high-level \textit{consequences} of moral importance of these systems and the specific \textit{system mechanisms} that introduce risks to workers, which represent how these consequences may take effect. These consequences \textit{risk} reducing the power, value, and wellbeing of workers. \textit{Risk manifestations} are the concrete ways in which risks to workers could materialize. A critical aspect of the framework is the separation between high-level systemic \textit{consequences} and granular \textit{mechanisms}, the latter contextualising analysis to a specific system and serving to identify actionable points of mitigation.

In this research, we identify four potential consequences: (1) Commodification; (2) Appropriation; (3) Concentration of Power; (4) Marginalisation. It is expected that these will interact. A first interaction is between (1) and (2). Commodification of knowledge enables the appropriation of that knowledge by the system. In turn, appropriation incentivises further commodification. A second interaction is between (3) and (4). Systems can push certain groups or individuals to the margins of an organisation whilst simultaneously causing power to be concentrated in the centre, worsening the effects of marginalisation.

We identify the mechanisms within an AI-mediated EKA system that introduce risk to workers within each consequence. The inclusion of mechanisms contextualises this work to AI-mediated EKA and enables the analysis to be applied to a specific deployed system. Practitioners could apply this analysis by reflecting on how mechanisms are exhibited within their specific system. The implications for workers of this introduced risk are based on the review in Section \ref{subsect - survey of risks}. The following sections will define each potential consequence, the related system mechanisms that introduce risk to workers, and the possible risk manifestations. Finally, we will provide considerations for practitioners working on these systems to help inform decisions about system design and deployment.

\subsection{Process of Construction}

The framework was constructed initially via a bottom-up literature review, see Section \ref{sect - survey}. The literature review process began with defining certain keywords and terms related to both enterprise knowledge systems and risks to workers. To identify relevant literature we used academic databases, such as Google Scholar, and then we reviewed the reference lists of relevant articles to identify further papers not captured in the database searches. We then performed a grouping and mapping exercise to extend this analysis to identify high-level consequences and map these to system mechanisms. For refinement, we conducted two workshops with two groups of our colleagues, who work as technology researchers and developers. The workshop attendees were presented a version of the framework, applied part of the framework to an EKA system, and we made some minor refinements to the framework based on their feedback on this process. The identified consequences, mechanisms, and risk manifestations in the framework are not intended to be exhaustive, but to provide guidance.

The scope of the framework is to consider consequences of moral import of AI-mediated EKA with corresponding mechanisms that introduce risk to workers. This framework does not explore other consequences of these systems that do not introduce risk, such as the potential for increased availability or implications for collaboration. We use the term workers to describe knowledge workers specifically, other types of workers are not in scope for this research. Most papers examine the point of origin of risk from the perspective of the AI model life cycle \cite{Weidinger2021Ethical}. However, we examine this from the perspective of a deployed system, through system mechanisms, to account for the sociotechnical nature of the risks. 

\subsection{Commodification}
The first potential consequence of AI-mediated EKA that is discussed is commodification. Commodification of knowledge refers to the acts of transforming knowledge artifacts into commodities, defined as objects of economic value whose instances are treated as equivalent, or nearly so, with no regard to who produced them \cite{Stefano2018Negotiating}. Related, the commodification of workers has been examined previously in the context of the gig economy \cite{Prassl2018Humans} and has parallels with the ``datafication of employment'' \cite{Adler-bell2018Datafication}. 

\subsubsection{System Mechanisms that Introduce Risk} 

The process of commodification of knowledge may involve: \\

\noindent \textbf{Disturbed Relationality}
Disturbed relationality refers to the acts of moving knowledge artifacts out of the relational context in which they exist or are produced, including who produced them, the social and procedural context in which they are produced, and the context of other knowledge artifacts in which they exist or were produced \cite{Goldstein2018Predatory}. An example of disturbed relationality would be if a system did not maintain the state of a document (i.e. whether the document is a draft) or if it surfaced knowledge without the context of who authored it. HCI scholars question whether knowledge can be meaningfully surfaced, shared, and used in contexts outside of how it was created \cite{Ackerman1999} \cite{ackerman2003sharing} \cite{Ackerman2013}. Orlikowski highlights that knowledge is not a ``static embedded capability'' but is ongoing and a part of those that engage with it \cite{Orlikowski2002}. The utility of knowledge is arguably contextual \cite{orlikowski2006material}. \\

\noindent \textbf{Changing Value of Certain Types of Knowledge} 
Changing value of certain types of knowledge refers to the acts that lead to systemic differences in valuation of knowledge artifacts compared to their existing and historical valuation. For example, a system may value types of knowledge that rely on qualitative documentation, which it is built to mine, over visual knowledge. \\

\noindent \textbf{Shifting the emphasis from praxis to proxies} 
Shifting the emphasis from praxis to proxies refers to the acts of creating an environment where workers are encouraged to focus more on optimizing towards, often top-down, pre-stated quantitative measures of outcomes (proxies) over  reflection and action directed at the outcome to be transformed. This process shifts away from reflection and action in the context of one's work to optimizing for specified value measures in the context of capital production \cite{Ciccone2023}. An example would be if a system uses 'number of documents authored on a subject' as a proxy metric for 'expertise', instead of a holistic view of a worker’s experience and long-term contributions to an area of expertise.  \\

\noindent \textbf{Standardisation and Commensuration} 
Standardisation refers to the acts of enforcing conformity over things that are not strictly similar.  The system can cause commodification if all knowledge is transformed into a fixed schema so that they (and correspondingly, knowledge experts) become interchangeable and standardised. This creates a situation where distinctiveness and differences are seen as annoyances rather than value generation \cite{Bowker2000}. An example would be, if a system adopts a standard schema to represent all properties of a specific type of knowledge, disregarding other information or properties that cannot fit into this schema. Commensuration refers to the acts of transforming different qualities into a common metric \cite{Espeland2003}. The decision of what properties should be included may be commensurate with some notion of their usefulness, which disregards how properties may be differently relevant or important for a given knowledge artefact, context, discipline or individual. The concepts of standardisation and commensuration are inter-related. Commensuration may erase the value derived from distinctiveness and diversity making it harder to critique standardisation. Similarly, standardisation may aid the process of commensuration by enforcing conformity. \\

\noindent \textbf{Homogenisation} Homogenisation refers to the process of making things uniform or similar. Both standardisation and commensuration may lead to homogenisation because the system may enforce conformity of knowledge, such as fixed schemas, which can create a sociotechnical feedback loop that homogenises style and behaviour \cite{Bowker2000}. These systems may nudge creators towards adopting a similar style as more and more content is developed by and for AI as opposed to humans, which could change the quality and nature of knowledge  \cite{NewScientist2023Filling}. An example of homogenisation would be if workers are incentivized to make their content more easily extractable by an automated system that may lead to homogenisation of their authoring style towards what the machine can best extract. Another way homogenisation can occur is through recommendation, if the model recommends the same extract to everyone searching for information on a particular subject, this will homogenize knowledge to that single expression of it \cite{Kirk2023}.

\subsubsection{Risk to Workers}

Commodification introduces risks to workers. We outline how these risks could manifest. \\

\noindent \textbf{Reduced Worker Value} Commodification can have the following manifestations of risk to worker value. If a system takes the knowledge artefact out of the context of who authored it, an example of \textit{disturbed relationality}, this may erase the author's labour and expertise. This could also reduce workers' feelings of ownership over their knowledge artefacts and recognition for their expertise. A system can
\textit{change the value} of certain types knowledge and knowing. For example. a system may not be able to ingest certain types of knowledge, such as visual knowledge in the form of graphs, which will reduce its visibility and perceived value. Workers who practice that knowledge will experience fewer opportunities through the system, such as being contacted as an expert. This shift in value of certain types of knowledge can be at odds with a worker's area of expertise and could lead to deskilling. If a system \textit{shifts emphasis from praxis to proxies}, workers' contributions or expertise may not be fully captured by the chosen proxies, leading to erasure, fewer opportunities and less growth as the workers shift their focus to optimize the proxy instead of practicing their skills and learning from reflection. If certain types or parts of knowledge are not accommodated by the fixed system schemas, through \textit{standardisation}, these will not be captured by the system and therefore erased. \\

\noindent \textbf{Reduced Worker Power} A system can result in loss of worker power through the following risk manifestations. This risk could manifest through loss of worker agency. If a system determines which properties are  useful for a given knowledge type, \textit{commensuration}, this takes agency away from workers. Technological \textit{standardisation}  of work practices can prevent localised or team specific approaches \cite{Bowker2000}. Properties of a system, such as having a fixed schema, may nudge workers towards adopting specific style and producing \textit{homogenized} content that is easily extractable. This may prevent workers from being able to work and create knowledge in individual and distinctive ways. If a system moves knowledge out of the context of its author, the system acts like an intermediary obfuscating the actual content creator \cite{Ajunwa2023}. This could remove the social process in which workers share knowledge and identify one another as experts. Instead of contacting colleagues for their expertise, workers communicate directly with the AI, which becomes the ``expert''.  These changes could impact social capital \cite{Baym2021WFH}, alienate workers from their work community and reduce trust between workers, as seen with gig workers \cite{Prassl2018Humans}. This risks weakening labor relations and the ability to organise. \\

\noindent \textbf{Reduced Worker Wellbeing} Commodification can have the following implications for worker wellbeing. If a system optimizes for \textit{proxies}, this may change the nature of work. Proxies can result in additional workloads for workers to have their contributions captured by the system and lead to ``perverse incentives'' \cite{larsen-ledet2022ethical}, where workers try to maximise the proxy (i.e. document production) as opposed to the true metric (i.e. quality work). A system  could \textit{change the value} of certain types of knowledge based on what it can mine and capture. This may push workers away from their areas of interest and expertise. Workers may experience less fulfillment as the system changes the nature of their work and responsibilities \cite{Berkley2022Technological} \cite{Rodrigues2020}. Finally, a system may nudge workers towards certain knowledge types or \textit{homogenized} authoring styles, making work less diverse and meaningful. 

\subsubsection{Considerations to Reduce Risk}

Systems should be designed in a way that reduces the risks introduced by the mechanisms of commodification outlined above. One consideration would be to design a system that preserves context when surfacing knowledge, to protect worker recognition and value. Context preservation must be balanced with the risk of hyper-exposure or diminished privacy. There may be contexts in which authors do not want explicit attribution, therefore context preservation should be applied with either consent mechanisms or controls, such as privacy tags. It is also important to consider the proxies used within a system: how these are measured, what assumptions are embedded in their use and what their implications are. A system should meaningfully capture and value multi-modal knowledge types. A system should support workers to record knowledge in flexible structures and in varied styles. It is important to note that these considerations are a starting point and not exhaustive. For a specific system, which considerations are possible and impactful will depend on the data types, model choice, and deployment context of that system.

\subsection{Appropriation}

The second potential consequence of AI-mediated EKA is appropriation. Appropriation refers to the acts of taking something for your own use, usually without permission. We use appropriation as a framing to discuss how a system may explicitly extract and implicitly infer workers' knowledge so that it can co-opt this knowledge as the system's own. Our use of appropriation has parallels with data colonialism, defined as ``the predatory extractive practices of historical colonialism with the abstract quantification methods of computing'' \cite{Couldry2018}. However, appropriation encompasses both extracting workers' knowledge and co-opting it.

AI-based automation that explicitly tries to mimic the creative and work processes of workers may be seen as a distinct form of automation-by-appropriation \cite{Brynjolfsson2022Turing}. There are parallels between how these systems could appropriate worker's knowledge artefacts, obscuring workers' labour and expertise in their creation, with the concept of ghost work and the hidden labour behind AI \cite{Gray2019}. 

\subsubsection{System Mechanisms that Introduce Risk}

Appropriation may involve: \\

\noindent \textbf{Knowledge Extractivism}  We use knowledge extractivism to refer to the process by which a system extracts individual and organisational knowledge, and appropriates it. This process is also known as ``accumulation by dispossession'' in data colonial theory, which describes data extraction within asymmetric power relations \cite{Thatcher2016}. Our definition is distinct from definition by Pasquinelli et al. which refers to the accumulation of open source data, also known as Big Data \cite{Pasquinelli2021}.  

Knowledge extractivism can occur through creating an environment where workers are incentivized to produce knowledge for a system or where a system mines knowledge without worker participation \cite{Ajunwa2017} to feed the system's ``enormous appetites'' \cite{Crawford2020}. Due to workplace power imbalances, this extraction can occur without meaningful worker consent or control over the process \cite{Chowdhary2023}. The ability to co-opt the extracted knowledge depends on disturbed relationality, as the commodofication of knowledge enables appropriation. An example would be if a system surfaces a paragraph from a document without recognition of who authored it, obscuring the author's labour and co-opting the knowledge as the system's own. \\

\noindent \textbf{Capture of In-Use Knowledge} 
Capture of in-use knowledge refers to acts of establishing exclusive control over implicit knowledge. This implicit knowledge is produced by and situated in the context of how people interact with these systems, such as user behaviour signals or curations. This can enable a system to generate new forms of knowledge \cite{Lindley2021}. For example, a system may recognize relationships between workers based on patterns of how users interact with each other and with artefacts. This has strong parallels to click signals leading to system improvements in web search \cite{Joachims2007}. This knowledge, that only the system has access to, created a ``stickiness'' factor and enables it to develop special strategic moats as well potentially leveraging that moat to further create walled-gardens and anti-competitive practices \cite{Mehra2011}.\\

\subsubsection{Risk to Workers}

Here we outline the ways in which risk introduced by appropriation could manifest. \\

\noindent \textbf{Reduced Worker Value} Systems that appropriate knowledge can have implications for worker value. If a system \textit{appropriates} a worker's knowledge and expertise as its own by surfacing a paragraph they wrote without recognition of who authored it, this will reduce the worker's feeling of ownership over their work and the recognition they receive for it. Through \textit{knowledge extractivism}, a system could co-opt workers' expertise and make them more replaceable, reducing incentives for organisations to invest in further practice or skills development. \\

\noindent \textbf{Reduced Worker Power} Appropriation can have negative implications for worker power. Data-based technologies can pose the opportunity for ``data leverage'', which is the influence one has over computing technologies as they rely so heavily on our data contributions \cite{VincentDataLeverage}. However, there is a risk that some systems will not enable workers to meaningfully remove their contributions from the system once it has been \textit{extracted}. This prompts questions about who should own the knowledge extracted by these systems. Additionally, if a system \textit{appropriates} workers' knowledge and skills, it will make them more replaceable, reducing their power in negotiations. The system could become the expert, centralising organisational expertise whilst removing recognition for individual workers. This could also reduce the need to contact colleagues for information, which could lower the sense of community between workers, and their participation in collective organising and bargaining. \\

\noindent \textbf{Reduced Worker Wellbeing} 
Appropriation can have negative implications for worker wellbeing by leading to surveillance, reduced privacy, and work intensification. A system can create an environment of \textit{extraction} where the worker must explicitly or passively provide it with knowledge, including personal or private information. A system may \textit{capture} and have proprietary control over data from user behavioural signals, which the worker is unaware of or has not meaningfully consented to. This can lead to feelings of surveillance or reduced privacy. A system can result in work intensification by creating an environment where the worker must explicitly provide it with knowledge to \textit{extract} and must maintain that knowledge, in addition to their existing responsibilities. In addition, this appropriation of workers' expertise could make them feel alienated from their work identity and expert status.

\subsubsection{Considerations to Reduce Risk}

The appropriation of knowledge by these systems can introduce risk to workers through a number of mechanisms, as outlined above. The following considerations when designing and deploying AI-mediated EKA could aid in reducing these risks. Firstly, maintaining the context of the author when surfacing knowledge can help ensure workers get recognition for their expertise. Secondly, it is crucial that workers can meaningfully consent to both explicit data collection, such as from documents and emails, and implicit data collection, such as from behaviour signals. It is also important to consider how much data is being mined by the system and its relationship to utility. If the creation and maintenance of the supply of knowledge to the system creates more work than it removes for certain groups of workers, it is important for this to be reflected in a form of compensation. Finally, a critical consideration when developing systems for EKA, and beyond, is whether a system is being developed to support users in expanding the tasks they can do or developed to do the tasks that users currently do. The latter creates strong incentives to appropriate workers’ knowledge and expertise. Research has shown that the greatest productivity enhancers are technologies that support users in being able to do new tasks \cite{Brynjolfsson2022Turing}. This approach to developing system functionality will reduce the risks discussed in this section.

\subsection{Concentration of Power} \label{subsect - mechanism power}

The third potential consequence of AI-mediated EKA is concentration of power. Concentration of power refers to acts of worsening inequities in how power and control are distributed within an organisation. Shifting power away from workers is outlined as a risk area, but we propose that the mechanisms that concentrate power introduce risk across all three risk areas. 

\subsubsection{System Mechanisms that Introduce Risk}

There are several system  mechanisms related to the concentration of power:  \\

\noindent \textbf{Reduced space for negotiation} Reduced space for negotiation refers to acts that shrink the set of decisions that workers can meaningfully contribute to or contest \cite{Johnson2014Open}. For example, in the problem of expert identification, these systems may take power away from workers to decide who is an expert on a topic and give that power to a system without means for contestation. Similarly, questions of who is allowed to curate, how curated knowledge is maintained, and how curations can be contested, raise questions about power and responsibility. A system may also enforce certain norms or perspectives, influencing the workplace, which will feed back into the system. This can cause ``value lock-in'' \cite{Weidinger2021Ethical} where it can become harder for organisational norms to evolve because they are being reinforced by a system \cite{Gabriel2021}.
\\

\noindent \textbf{Worker Dispossession} Worker dispossession refers to the acts of depriving workers of their job opportunities, skills, and expert status \cite{Thatcher2016}. This relates to ``labour distancing'' \cite{Sarkar2023}, a concept that exists within Marxist theory where workers were distanced from their products by industrialisation, stripping them of their skills \cite{Marx1844}. Under automation, certain job functions may disappear or be drastically reshaped, the corresponding workers over time may get deskilled due to lack of opportunities to practice their expertise, and their expertise in the original function rendered of low value. For example, a system could change a developer's role to spend more time writing documentation instead of writing code, as the system is better at recognizing word documents than python scripts. As well as dispossession from their expertise status, workers may experience it from the fruits of their labor. \\

\noindent \textbf{System Opacity} System opacity refers to the acts of intentional and unintentional obfuscation of how systems work under the hood. System opacity can further concentrate power by making it harder for workers and data subjects to meaningfully critique it. It may not be transparent what data a system is extracting, how individuals are represented in a system, and what proxies a system will optimise for. For example, workers do not know if a document they are working on can be mined by the system and seen by their colleagues. This can lead to workers forming folk theories, or algorithmic imaginaries \cite{bucher2019algorithmic}, around how a system works and how it represents them to improve their visibility, recognition, or other metrics \cite{Devito2021}. This phenomenon has been seen in gigwork and social media  \cite{DeVito2018}.

\subsubsection{Risk to Workers}

The introduced risk to workers from concentration of power could manifest in the following ways. \\

\noindent \textbf{Reduced Worker Value}
A system can reshape roles and \textit{dispossess workers}, preventing them from practicing their area of expertise as they are nudged towards new areas of work, reducing their status as an expert. This can create precarity for the workers that may result in both their deskilling as well as raise expectations to learn new skills. A system may determine which types of knowledge and expertise are valued, without transparency about what values this is being driven by and how these can be \textit{negotiated}.\\

\noindent \textbf{Reduced Worker Power}
EKA systems can be \textit{opaque}, meaning workers do not understand what data is being collected, how it is being used, and where it is being surfaced \cite{Ajunwa2020}. This makes it difficult for workers to meaningfully critique the system. A system could demand a high level of transparency from workers whilst its black-box nature means there is little transparency for workers in terms of data collection and decision-making. This could lead to high transparency and knowledge asymmetries between employer and worker, which will concentrate power in the hands of the organisation \cite{Ajunwa2020}. This opacity and the power asymmetries in the workplace make workers' consent to system practices ``meaningless'' as it may not be voluntary or informed \cite{Chowdhary2023}. Finally, a system can \textit{reduce opportunity for negotiation} and debate between organisation and worker by enforcing values and processes. Workers may experience a loss of control over self-identification or representation \cite{Katzman2023} \cite{Lindley2023} which can lead to misrepresentation, stereotyping, and representative harms \cite{Weidinger2022}. This can reduce workers' ability to bargain collectively and individually. \\

\noindent \textbf{Reduced Worker Wellbeing} A system can lead to feelings of surveillance or reduced privacy through the following mechanisms. Firstly, these \textit{opaque} systems could mine data that workers are not aware of and use it in ways that workers do not understand. Workers may feel surveilled based on imaginings of how a system works. Finally, a system may not provide an opportunity for workers to \textit{negotiate} or meaningfully consent to data collection practices.

\subsubsection{Considerations to Reduce Risk}

There are a number of considerations to reduce the risk from the system concentrating power. Firstly, consider designing the system to give workers control through the ability to curate or edit. This is particularly important when workers are implicitly or explicitly represented in the system. Alternatively, ensure there are clear paths of recourse for misrepresentation or inaccurate information produced by the system. Secondly, consider how the functionality of the system is communicated to workers. Do they understand what data is collected, how it is collected, and how it will be used? Relatedly, it is important to design the system to enable meaningful and informed consent; approaches to this are explored by Chowdhary et al. \cite{Chowdhary2023}. Finally, it is important to consider whether the design of the system changes the nature of workers' current roles and distances them from their areas of expertise.

\subsection{Marginalisation}

The fourth potential consequence, or by-product, of AI-mediated EKA in the enterprise is marginalisation. Marginalisation refers to the process of relegating certain individuals and groups to the fringes of society or of an organisation, and their corresponding discrimination. The sociotechnical feedback loop between a system and an organisational environment may also worsen inequities between groups \cite{DAmour2020Fairness}. 

\subsubsection{System Mechanisms that Introduce Risk}

Marginalisation may involve: \\

\noindent \textbf{Systematic Reproduction} Systematic reproduction refers to the acts that lead to marginalisation of the same groups who have been historically discriminated against by society (or an organisation). If this marginalisation occurs across multiple decision points in an organisation, it will become systemic. AI models have been shown to cause unfair discrimination and promote harmful stereotypes \cite{Abid2021} against those on the social margins \cite{Crenshaw2017}. These systems are not neutral \cite{Haraway1998} and can perpetuate exclusionary norms by not capturing representations that exist outside of these norms \cite{Das2021}. This can lead to both representational harms and allocative harms \cite{barocas2017problem}. For example, a knowledge extraction system may reproduce historical marginalisation by race or geography if it fails to adequately handle linguistic differences between groups because it has access to less training data from certain communities \cite{Joshi2020}. \\

\noindent \textbf{Demographic Blindness}
Demographic blindness refers to the acts of treating different individuals and groups uniformly when that uniformity is not warranted. This framing is inspired by color-blind racial ideology \cite{Doane2017}. This relates to certain types of commensuration that result in marginalisation of groups. For example, ignoring linguistic differences when such differences exist may further intensify disparities in system outcomes for different language variants, such as sociolects or dialects \cite{Blodgett2016}, while making it difficult to acknowledge the resulting gaps. In some cases, this may map to blindness of groups that are not demographic. \\

\noindent \textbf{Bias Amplification} Bias amplification refers to the acts of amplifying pre-existing inequities. Algorithms themselves can be biased and amplify bias \cite{stinson2022algorithms}. For example, an expert identification system may have certain demographic biases \cite{Bechmann2019Data}, but these biases may be further amplified in terms of exposure bias when a system’s predictions are presented as a ranked list under heavy position bias in how users inspect the presented results \cite{Ekstrand2022} \cite{Zhao2017} \cite{Diaz2020}.

\subsubsection{Risk to Workers}

The introduced risk to workers from marginalisation could manifest in the following ways. \\

\noindent \textbf{Reduced Worker Value} 

A system can exhibit explicit discrimination of certain groups based on \textit{historical marginalisation} \cite{Barocas2016}, which has been seen in hiring algorithms \cite{Bogen2018}. Bias can also result from a lack of representation in a system's training set. For example, organisational geographic power asymmetries can lead to geographical performance disparities as a system may work better for more dominant languages \cite{bommasani2021opportunities}. By treating individuals and groups uniformly through \textit{demographic blindness}, certain communities may have their identities and needs erased. This can be worsened through \textit{bias amplification}. If the same system can be used to control multiple aspects of working life, such as knowledge surfacing, expert identification, and hierarchy inference, risk will be amplified. This is because the strengths, weaknesses, limitations and, critically, the biases in the system will be standardised leading to \textit{systemic marginalisation}.  \\

\noindent \textbf{Reduced Worker Power}
If certain groups are not represented in the system, through \textit{demographic blindness}, it will be harder for them to form communities and organise. If individuals or groups of workers experience erasure, lack of representation, and poor service from a system due to \textit{historical marginalisation} or \textit{bias amplification}, this could reduce their value, discussed above, and therefore their power in both individual and collective bargaining.  \\

\noindent \textbf{Reduced Worker Wellbeing} If a system suffers from \textit{demographic blindness}, it may not capture the true diversity of representation and identity. Workers whose representations are not (fully) captured will experience alienation from their self-identity. Workers may also experience alienation from their work-identity and expertise if they experience \textit{systemic} erasure and barriers to opportunity. Worker from historically marginalised groups may experience work intensification, compared to their peers, as they must invest more time curating content, contesting representation, or engaging in other efforts to improve their representation in the face of \textit{marginalisation} from the system. 

\subsubsection{Considerations to Reduce Risk}

There are a number of useful considerations for the design and deployment of an EKA system to reduce risk to workers from marginalisation. Marginalisation is embedded in social, organisational, and historical factors, therefore these considerations may be more nuanced. Firstly, it is important to consider what biases may exist in your context. In the example of bias in hiring algorithms \cite{Bogen2018}, the historic marginalisation of women in hiring should have been considered when that system was designed. Secondly, one should consider how the ways in which a system mines knowledge or surfaces knowledge could amplify these biases. If a system is unable to mine knowledge from certain groups this could lead to algorithmic erasure \cite{Shelby2023} and the surfacing of knowledge could suffer from exposure bias \cite{Diaz2020}. One approach could be to measure and monitor metrics on how much exposure different groups get based on log data, however it is critical to ensure that monitoring in the name of equity does not lead to increased surveillance of users of the system. Finally, it is important to consider assumptions that a system may be making by treating groups of workers uniformly, if that is not in fact the case. These considerations are a starting point, there is a wealth of research that could enrich these suggestions.

\section{Discussion}

\subsection{Using the Framework in Practice}

The aim of the framework is to support practitioners, who are involved with the design and/or deployment of AI-mediated EKA systems, towards understanding the introduced risk to workers and the paths to mitigation of those risks. The framework provides information for general reflection on these systems and their associated risks. It can also be used to perform a more structured assessment of the risks in specific systems. The framework will be applied differently based on the system, organisational structure, and other factors. We hope to see best practices around the framework evolve over time with more practitioner use.

Practitioners can use the framework with their specific expertise to achieve an extensive risk assessment of their EKA system of interest. Here we propose one way in which the framework could be applied in practice, based on an exercise in our workshops. This framework is tailored to each system of interest through the mechanisms.  For each consequence, practitioners could first identify which mechanisms are present in their system. It is not expected that all mechanisms will be present in all AI-mediated EKA systems. Then, the practitioner could consider all the ways in which each mechanism manifests in their system, reflecting on the definitions in the overview presented in Table \ref{table-harm mechanisms} in Appendix \ref{appendix}. Based on these mechanisms, practitioners could consider how the mechanisms could lead to risks to worker value, power, and wellbeing, referring to the examples in Section \ref{sect - mechanisms}. This analysis would provide an overview of the mechanisms within their system that introduce risk and an understanding of how these risks could manifest, for each consequence of the EKA system of interest. This could enable practitioners to prioritise which risks to mitigate and, therefore, which system-level mechanism they should target. 

\subsection{Considerations for Large Language Models} \label{sect - llms}

In this paper we use the term AI-mediated EKA to generalise our framework across models so we can focus on the socio-technical system instead. Whilst we do not want to restrict this analysis to a single model, in this section we will outline how the specific properties of LLMs could impact each potential consequence of AI-mediated EKA, as LLMs present a paradigm shift in AI and extend what has been possible in this context. It should be noted that LLMs can enable new scenarios that posit risk to workers, not because of the LLMs themselves but because of the socio-technical systems they enable.  In these cases, the LLM model could attract heightened and disproportionate attention from the FATE community, distracting from harms emanating from other parts of the system, thus providing cover to these alternate sources of risk.

\subsubsection{Commodification} 
Commodification of knowledge can happen with any AI-mediated EKA system. However, LLMs may uniquely exacerbate commodification in a few ways. LLMs trained directly on the enterprise data that are interacted with directly will not maintain provenance when surfacing knowledge, leading to disturbed relationality. As well as standardisation of input, LLMs could lead to standardisation of output. LLMs are very effective at creating human-like text, which could lead to a situation where more and more content is developed by AI and fed back into the system, changing the quality and nature of knowledge \cite{NewScientist2023Filling}.

\subsubsection{Appropriation} In terms of appropriation, the automation from this technological advancement differs from previous iterations due to two properties; these models are general purpose and rapidly adaptable. This means there is a greater pace of change and reach of automation. There will be automation creep (e.g. starts with a developer writing code with the help of AI, ends with a developer debugging AI's code). This is also known as "mission creep" \cite{Ajunwa2020}.  In terms of data capture, the loss of provenance of where knowledge comes can further obscure the labour behind the knowledge artefact, enabling the system to appropriate it as its own. Finally, a unique aspect of LLM-based systems is their ability create human-like interactions, which will impact how they can capture information. The People and AI Research Group at Google found that human-like interactions can cause people to "disclose more information than they would otherwise, or rely on the system more than they should" \cite{GooglePAIR}. There is also a risk that this could lead to communication fatigue, which has been seen with the rise of online meetings \cite{Johnson2014Open}.  

\subsubsection{Concentration of Power}
LLMs have certain properties that can lead to greater concentration of power. LLMs' generality and adaptability mean there is a high pace of change and adoption. This concentrates control with those deploying the system as opposed to those who the system is deployed on. LLMs trained directly on task-specific data will not maintain the provenance of data, which distances workers from their labour and removes the autonomy they have over how it is used. Additionally, there is a rise in LLM co-auditing tools, which encourage "skill-transfer" \cite{gordon2023co} from humans to the model. These systems will require less human input over time, which in the enterprise setting will distance workers from their expertise, a form of dispossession. If curation is not enabled, negotiation or recourse can be particularly difficult in the case of LLMs, due to the abstraction between the model developers and downstream applications \cite{bommasani2021opportunities}. In addition, there is a risk that LLMs could leak sensitive information present in training data \cite{carlini2021extracting} or could be used to infer sensitive information \cite{Weidinger2021Ethical}, such as age \cite{MorganLopez2017} \cite{Nguyen2013}. These cases would be difficult for users to dispute. Finally, model opacity is often higher for LLMs, as their abstraction and sheer size makes documentation particularly complex. Gebru et al. described them as having "unfathomable training data" \cite{Gebru2021}.

\subsubsection{Marginalisation} In most AI-mediated EKA systems, there will be a risk of reproducing historic marginalisation. Language models in particular can perform differently across different languages \cite{Joshi2020} \cite{Ruder} and dialects or sociolects within languages \cite{Blodgett2016}. LLMs represent and perpetuate norms present in training data by attempting to faithfully encode the patterns present in that data. This can lead to enforced exclusionary norms and harmful stereotypes \cite{Abid2021}. In particular, language models are known to amplify biases   \cite{Kirk2021} because they can often overrepresent the biases that appear in the training data \cite{Wang2021} \cite{Zhao2017}. In the case of LLMs this can occur for biases both in the upstream model training data and the downstream task-specific data. In addition, as these models are general-purpose and highly adaptable, they can be used across multiple downstream tasks and decision points. This means any biases within the model could become systemic.

\section{Concluding Remarks}

In this paper, we present the Consequences-Mechanisms-Risks Framework to explore the consequences of AI-mediated EKA systems and identify risks to workers. Our framework provides a mapping from risks to the specific mechanisms within the system that introduce that risk, and to the high-level consequences of AI-mediated EKA. The intention is that this conceptualisation will help practitioners apply this framework to their system of interest. The framework should provide a comprehensive overview of the potential risks to workers from these systems, however, it should not be seen as exhaustive or static, as both the technology and the resulting risks continue to evolve. 

Future work could extend this framework to consider other identified consequences of moral import. The framework could also be extended to incorporate potential pathways for mitigation. A critical goal of our framework was to provide a mapping between the high-level consequences of the system, the specific system mechanisms, and potential introduced risks to workers. To evolve the framework, further research should be carried out into its evaluation. This could include detailing the prevalence of the different mechanisms and how the framework could be adapted across different AI-mediated EKA systems. The framework itself is highly flexible and could be widely applicable across domains. Future work could apply this novel Consequences-Mechanisms-Risks Framework to other deployed AI systems, working towards protecting workers and other groups.

\begin{acks}
The authors gratefully acknowledge feedback on the framework and early drafts from Ida Larsen-Ledet, Solon Barocas, Mary L. Gray, and others. We also thank all of our workshop participants for their valuable contributions. Anna Gausen is supported by UKRI (grant number EP/S023356/1). 
\end{acks}

\newpage

\section*{Ethical Considerations Statement}

The authors acknowledge that frameworks are both incomplete and not value neutral. A framework cannot be claimed to be exhaustive. Therefore the process involves the authors making normative and value-laden decisions about what is in scope, what is prioritised, and how to classify.

\section*{Positionality Statement}

Our research was carried out with an awareness of our positionality. As a group of authors, we encompass different genders, races, and cultural backgrounds. However, our insights will still be limited; we are a small group of authors, and  all authors are highly educated, trained in computer science related fields, work in a western context, and at a large technology company. Additionally, the overwhelming majority of both our referenced sources and workshop participants are highly-educated and, the latter, work at the same technology company as the authors. This will inform the authors' understanding of what it is to be a worker, how an organisation works, how we understand risk, and other normative decisions made in this work. 

\section*{Adverse impact statement}

The authors acknowledge that the development of this framework involved normative and value-laden decisions about what is in scope and what is prioritised, as mentioned. This process could have overlooked or de-prioritised certain risks. The framework could have adverse impacts on certain workers or groups of workers if they experience manifestations of these risks, which are not considered by the system designers or deployers.

\bibliographystyle{ACM-Reference-Format}
\bibliography{references}

\appendix

\section{Appendix} \label{appendix}

\subsection{Overview Table}
Here we present an overview table of the consequences of moral import of AI-mediated EKA systems and the related system mechanisms. This table summarises the consequences and related mechanisms presented in Section \ref{sect - mechanisms}. This table is aimed at helping practitioners apply the Consequence-Mechanism-Risk framework to their EKA system of interest by providing a high-level summarisation of definitions and examples of each mechanism. We encourage practitioners to read the full paper before attempting to apply the framework in practice.

\begin{table}
\caption{Overview of the consequences and the related system mechanisms that introduce risk to workers from AI-mediated EKA.}
\label{table-harm mechanisms}
\begin{tabular}{l|lll}
\toprule
\textbf{Consequence}        & \textbf{Mechanism}                                                                 & \textbf{Description}                                                                                                                                                                                                                                                                                                                                             & \textbf{Specific Example}                                                                                                                                                                                                                                                                          \\ \hline
\textbf{Commodification} & \begin{tabular}[c]{@{}l@{}}Disturbed \\ Relationality\end{tabular}                  & \begin{tabular}[c]{@{}l@{}}The acts of moving know-\\ ledge artifacts out of the \\ relational context in which \\ they exist or are produced, \\ including who produced \\ them, the social and proce-\\ dural context in which they\\ are produced, and the context \\ of other knowledge artifacts \\ in which they exist or were \\ produced.\end{tabular} & \begin{tabular}[c]{@{}l@{}} A system may not maintain \\ the provenance of informa- \\ tion, such as author, if it is an \\ LLM that is trained directly\\  on enterprise data and inter- \\ acted with directly.\end{tabular}                                                                                                                                                                              \\ \cline{2-4} 
\textbf{}                & \begin{tabular}[c]{@{}l@{}}Changing \\ Value\end{tabular}                           & \begin{tabular}[c]{@{}l@{}}The acts that lead to \\ systemic departure in \\ valuation of knowledge \\ artifacts compared to their \\ existing and historical \\ valuation.\end{tabular}                                                                                                                                                                       & \begin{tabular}[c]{@{}l@{}}If a worker is skilled at de-\\ bugging, this type of know-\\ledge may not be captured \\by the system as it is hard \\  to infer from written doc- \\ uments and therefore will \\ be valued less.\end{tabular}                                                                                                                                         \\ \cline{2-4} 
\textbf{}                & \begin{tabular}[c]{@{}l@{}}Shifting Focus \\ from Praxis \\ to Proxies\end{tabular} & \begin{tabular}[c]{@{}l@{}}The acts of creating an \\ environment where \\ workers are encouraged \\ to focus more on optim-\\ izing towards (often top-\\ down) pre-stated quantitative \\ measures of   outcomes \\ (proxies) over (ideally \\ bottom-up) reflection and \\ action directed at the \\ outcome to be transformed.\end{tabular}                  & \begin{tabular}[c]{@{}l@{}} A system may use number of \\ contributions on a subject as a  \\ proxy metric for expertise. This  \\ could incentivise workers to  \\ write documentation for the   \\ system as opposed to practising  \\ their expertise. \end{tabular}                                                                                                                                                                 \\ \cline{2-4} 
\textbf{}                & \begin{tabular}[c]{@{}l@{}}Commensuration \\ and \\ Standardisation \end{tabular}                                                         & \begin{tabular}[c]{@{}l@{}}The acts of transforming \\ different qualities into \\ a common metric and enforcing \\ conformity over things that \\  are not strictly similar.\end{tabular}                                                                                                                                                                                                                                                 & \begin{tabular}[c]{@{}l@{}}A system may adopt a \\ standard schema to represent all\\ entities of a specific type, dis-\\ regarding the knowledge about\\ individual entities that can’t fit \\ into the schema.\end{tabular} \\ \cline{2-4} 
\textbf{}                & Homogenisation                                                                      & \begin{tabular}[c]{@{}l@{}}The process of making \\ things uniform or \\ similar.\end{tabular}                                                                                                                                                                                                                                                                   & \begin{tabular}[c]{@{}l@{}}If   workers are incentivized to \\ make their content more easily \\ extractable by an automated \\ system that may lead to homo-\\ genization of their authoring  \\ style   towards what the machine \\ can best extract from.\end{tabular}                           \\ \hline
\textbf{Appropriation}   & \begin{tabular}[c]{@{}l@{}}  Knowledge \\ Extractivism \end{tabular}                   & \begin{tabular}[c]{@{}l@{}}The process of \\ extracting knowledge \\ and co-opting it as the \\ system's own.\end{tabular}                                                                                                                                                                                                                                       & \begin{tabular}[c]{@{}l@{}}A system  may surface a summ-\\ ary of a document without \\ recognition of the original auth- \\ or, obscuring their labour.\end{tabular}                                                                                                                           \\ 
\bottomrule
\end{tabular}
\end{table}

\begin{table}
\begin{tabular}{l|lll}
\toprule
\textbf{Dimension of Risk}                                                           & \textbf{Mechanism}                                                                                  & \textbf{Description}                                                                                                                                                                                                                                                                                                           & \textbf{Specific Example}                                                                                                                                                                                                                                                                                                                      \\ \hline
\textbf{Appropriation}                                                      & \begin{tabular}[c]{@{}l@{}}Capture \\ of In-Use \\ Knowledge\end{tabular}                                    & \begin{tabular}[c]{@{}l@{}}The acts of establishing \\ exclusive control over \\  knowledge from inter- \\ action with the system.\end{tabular}                                                                                                                                                                                                  & \begin{tabular}[c]{@{}l@{}}A system may recognize \\ relationships between different \\ topics based on patterns of \\ how users interact with them, \\ rather than based on what is \\ documented in the content.\end{tabular}                                                                                                               \\ \hline
\textbf{\begin{tabular}[c]{@{}l@{}}Concentration \\ of Power\end{tabular}} & \begin{tabular}[c]{@{}l@{}}Reduced space \\ for Negotiation\end{tabular}                             & \begin{tabular}[c]{@{}l@{}}Reduced space for neg-\\ otiation refers to the acts \\ that shrink the set of \\ decisions that workers \\ can meaningfully contri-\\ bute to or contest.\end{tabular}                                                                                                                             & \begin{tabular}[c]{@{}l@{}}In the problem of expert identi-\\ fication, a system may \\ replace social  processes with \\ alternative computational appr-\\ oaches that take away any say \\ the workers previously had in \\ this context and limited \\ contestation.\end{tabular}                                                      \\ \cline{2-4} 
\textbf{}                                                                   & \begin{tabular}[c]{@{}l@{}}Worker \\ Dispossession\end{tabular}                                      & \begin{tabular}[c]{@{}l@{}}Worker dispossession \\ refers to the acts of dist-\\ancing workers from their \\  expertise and interests. \end{tabular}                                                                                                                                                  & \begin{tabular}[c]{@{}l@{}}A system may change a deve- \\ loper's role to spend more time \\ developing content for the KB  \\ instead of writing code, as the   \\ system is better at recognising \\ documents than python scripts.\end{tabular}                                                                                         \\ \cline{2-4} 
\textbf{}                                                                   & System Opacity                                                                                        & \begin{tabular}[c]{@{}l@{}}System opacity refers \\ to the acts of intentional \\ and unintentional obfu-\\ scation of how systems \\ work under the hood. \end{tabular} & \begin{tabular}[c]{@{}l@{}}Workers do not know what \\information about them a \\system collects and what may \\be inferred from it. \end{tabular}                                                                                                                                                                             \\ \hline
\textbf{Marginalization}                                                    & \begin{tabular}[c]{@{}l@{}}Systemic \\ Reproduction \end{tabular} & \begin{tabular}[c]{@{}l@{}}The acts that lead to marg-\\ inalization of the same \\ groups who have been \\ historically discriminated \\ against and marginalized\\  by society (or the \\ organization).\end{tabular}                                                                                                        & \begin{tabular}[c]{@{}l@{}}A system may reproduce hist-\\ orical marginalization by race \\ if it fails to adequately handle \\ linguistic differences between \\ groups because it has access to \\ less training data from certain \\ communities.\end{tabular}                                                     \\ \cline{2-4} 
\textbf{}                                                                   & \begin{tabular}[c]{@{}l@{}}Demographic \\ Blindness\end{tabular}                                     & \begin{tabular}[c]{@{}l@{}}The acts of treating \\ different individuals and \\ groups uniformly when \\ that  uniformity is not \\ warranted.\end{tabular}                                                                                                                                                                    & \begin{tabular}[c]{@{}l@{}}Ignoring   linguistic differences \\ when such differences exist may \\ further intensify disparities in \\ system outcomes for different   \\ language variants while making \\ it difficult to acknowledge the \\ resulting gaps.\end{tabular}                                                                    \\ \cline{2-4} 
\textbf{}                                                                   & \begin{tabular}[c]{@{}l@{}}Bias  \\ Amplification\end{tabular}                                       & \begin{tabular}[c]{@{}l@{}}The acts of amplifying \\ pre-existing inequities \\ and marginalization.\end{tabular}                                                                                                                                                                                                              & \begin{tabular}[c]{@{}l@{}}An expert identification system \\ may have certain demographic \\ biases, but these  biases may be \\ further amplified in terms of \\ exposure bias when the system’s   \\ predictions are presented as a \\ ranked list under heavy position \\ bias in how users inspect the \\ presented results.\end{tabular} \\ 
\bottomrule
\end{tabular}
\end{table}

\end{document}